\documentclass{emulateapj}
\def \be{\begin{equation}}
\def \ee{\end{equation}}
\def \bdm{\begin{eqnarray}}
\def \edm{\end{eqnarray}}
\begin{document}
\title{The implicit contribution of slab modes to the perpendicular diffusion coefficient of particles interacting with two-component turbulence}
\author{A. Shalchi}
\affil{Department of Physics and Astronomy, University of Manitoba, Winnipeg, Manitoba R3T 2N2, Canada}
\email{andreasm4@yahoo.com}
\begin{abstract}
We explore the transport of energetic particles in two-component turbulence in which the stochastic magnetic field
is assumed to be a superposition of slab and two-dimensional modes. It is known that in magnetostatic slab turbulence,
the motion of particles across the mean magnetic field is subdiffusive. If a two-dimensional component is added,
diffusion is recovered. It was also shown before that in two-component turbulence, the slab modes do not explicitly contribute
to the perpendicular diffusion coefficient. In the current paper the implicit contribution of slab modes is explored and
it is shown that this contribution leads to a reduction of the perpendicular diffusion coefficient. This effect improves
the agreement between simulations and analytical theory. Furthermore, the obtained results are relevant for investigations
of diffusive shock acceleration.
\end{abstract}
\keywords{diffusion -- magnetic fields -- turbulence}%
\section{Introduction}
The interaction between energetic particles and a magnetized plasma is explored analytically. Examples for energetic
particles are Solar Energetic Particles (SEPs) and Cosmic Rays (CRs). If such particles move through space, their motion
is usually diffusive and, therefore, a diffusive transport equation has to be used in order to describe their motion.
The most important terms in such transport equations are those describing spatial diffusion along and across a mean
magnetic field. Parallel and perpendicular diffusion coefficients are mostly controlled by the turbulent magnetic fields
of the plasma.

First treatments of perpendicular diffusion were based on quasi-linear theory (see Jokipii 1966) which can be understood
as a first-order perturbation theory. However, perturbation theory is usually based on the assumption that there is a small
parameter. It is often stated in the literature (see, e.g., Schlickeiser 2002) that this small parameter is the magnetic
field ratio $\delta B / B_0$ (here $\delta B$ is the total turbulent field and $B_0$ is the mean magnetic field). Apart from
the problem that this magnetic field ratio is usually not small in astrophysical scenarios, it was shown in the literature
that a small value of $\delta B / B_0$ alone does not justify the quasi-linear approach (see, e.g., Shalchi 2009 for a detailed
discussion of the problems associated with quasi-linear theory).

Because quasi-linear theory is problematic, non-linear theories have been developed mainly in order to describe perpendicular
diffusion. Some work was already presented in the seventies of the 20th century (see, e.g., Owens 1974). A breakthrough has been
achieved by Matthaeus et al. (2003) where the so-called {\it Non-Linear Guiding Center (NLGC) theory} was presented. The latter
theory agrees well with simulations for a specific turbulence model, namely a so-called two-component model in which it is assumed
that the turbulence can be approximated by a superposition of slab and two-dimensional modes. However, NLGC theory does often
not provide the correct result. This is in particular the case for slab turbulence, two-component turbulence with a dominant
slab contribution, or three-dimensional turbulence with small Kubo numbers\footnote{The Kubo number occurs in investigations
of three-dimensional turbulence and is defined as $K = (l_{\parallel} \delta B_x)/(l_{\perp} B_0)$. Here we have used characteristic
lengths scales describing the correlation of the turbulent fields in the directions parallel and perpendicular with respect to the
mean magnetic field. Furthermore, $\delta B_x$ is the $x$-component of the turbulent magnetic field vector and $B_0$ is the mean field.}
(see, e.g., Shalchi 2006, Tautz \& Shalchi 2011, and Shalchi \& Hussein 2014).

In Shalchi (2010) the {\it Unified Non-Linear Transport (UNLT) theory} was developed. Although the theory is based on some of
the approximations used by Matthaeus et al. (2003), it contains a very different treatment of higher order correlations.
UNLT theory uses an approach based on the CR Fokker-Planck equation in order to avoid simple approximations of such
correlations. UNLT theory provides a non-linear integral equation similar compared to the NLGC result but it contains
different terms in the denominator. In particular for slab and small Kubo number turbulence, NLGC and UNLT theories provide
completely different results. Furthermore, UNLT theory contains the Matthaeus al. (1995) theory for field line random
walk as special limit.

UNLT theory provides the correct subdiffusive result for perpendicular transport in slab turbulence. Furthermore, the theory
states that the slab contribution in two-component turbulence damps out subdiffusively even if a dominant two-dimensional
component is added (see also Shalchi 2005 and Shalchi 2006). Therefore, slab modes do not explicitly contribute to the
perpendicular diffusion coefficient. It is the purpose of the current paper to explore the implicit contribution of the
slab modes to the perpendicular diffusion coefficient.

The paper is organized as follows. In Sect. 2 we discuss different analytical theories for perpendicular diffusion.
In Sect. 3 we developed a non-linear diffusion theory for two-component turbulence which takes into account the implicit
contribution of the slab modes. In Sect. 4 we present some analytical approximations which are useful in order to simplify
the new integral equation and in Sect. 5 we show the perpendicular diffusion coefficients based on different theories and
compare them with each other. In Sect. 6 we summarize and conclude. Furthermore, we point out which theory should be applied
for two-component and three-dimensional turbulence, respectively.
\section{Different analytical theories for perpendicular transport}
In the current section we briefly discuss three non-linear theories for perpendicular diffusion developed in the past.
Those are the original NLGC theory of Matthaeus et al. (2003), the Extended Non-Linear Guiding Center (ENLGC) theory of
Shalchi (2006), and the UNLT theory of Shalchi (2010).
\subsection{The original NLGC theory}
The original NLGC theory was developed based on different assumptions and approximations. In the following we briefly re-derive
this theory. This is necessary to point out the differences between different theories but some of the assumptions and
approximations used here will be employed in Sect. 3 in order to achieve a further improvement of the analytical description
of perpendicular diffusion.

As starting point we can use the following equation of motion (see, e.g., Schlickeiser 2002)
\be
v_x (t) = v_z (t) \frac{\delta B_x \left[ \vec{x} (t) \right]}{B_0}.
\label{eqofmotion}
\ee
Strictly speaking, the velocity component $v_x$ used here is the corresponding component of the guiding center velocity.
Matthaeus et al. (2003) introduced a {\it correction parameter $a$} in Eq. (\ref{eqofmotion}). In recent numerical investigations,
however, it was shown that $a = 1$ (see Qin \& Shalchi 2016). A more detailed discussion of this matter can be found below.

A diffusion coefficient can be calculated by employing the {\it Taylor-Green-Kubo formula} (see Taylor (1922), Green (1951),
and Kubo (1957))
\be
\kappa_{\perp} = \int_{0}^{\infty} d t \; \left< v_x (t) v_x (0) \right>
\ee
and with Eq. (\ref{eqofmotion}) we obtain
\be
\kappa_{\perp} = \frac{1}{B_0^2} \int_{0}^{\infty} d t \; \left< v_z (t) v_z (0) \delta B_x \left[ \vec{x} (t) \right] \delta B_x \left[ \vec{x} (0) \right] \right>.
\label{kappawith4th}
\ee
To continue, Matthaeus et al. (2003) have employed the following approximation
\bdm
& & \left< v_z (t) v_z (0) \delta B_x \left[ \vec{x} \left( t \right) \right] \delta B_x^{*} \left[ \vec{x} \left( 0 \right) \right] \right> \nonumber\\
& \approx & \left< v_z (t) v_z (0) \right> \left< \delta B_x \left[ \vec{x} \left( t \right) \right] \delta B_x^{*} \left[ \vec{x} \left( 0 \right) \right] \right>.
\label{4thorderapprox}
\edm
It was shown analytically in several papers (see, e.g., Shalchi 2006 and Shalchi 2010) that this type of approximation does not work for slab or slab-like
turbulence. Recent numerical tests have shown that this type of approximation works well for two-dimensional dominated turbulence but fails completely
for slab dominated turbulence (see Qin \& Shalchi 2016).

If approximation (\ref{4thorderapprox}) is combined with Eq. (\ref{kappawith4th}), we derive
\be
\kappa_{\perp} = \frac{1}{B_0^2} \int_{0}^{\infty} d t \; \left< v_z (t) v_z (0) \right>
\left< \delta B_x \left[ \vec{x} (t) \right] \delta B_x \left[ \vec{x} (0) \right] \right>.
\label{peprwith4thorder}
\ee
For the parallel velocity correlation function we employ an isotropic exponential model\footnote{A detailed investigation of velocity
correlation functions was presented in Shalchi (2011). It was shown there, that only if the pitch-angle Fokker-Planck coefficient is
isotropic $D_{\mu\mu} = D (1-\mu^2)$, we indeed find an exponential velocity correlation function. In other cases, however, it is much
more complicated to determine the analytical form of $\langle v_z (t) v_z (0) \rangle$.}
\be
\left< v_z (t) v_z (0) \right> = \frac{v^2}{3} e^{-v t / \lambda_{\parallel}}.
\label{isovelo}
\ee
To model the magnetic field correlations is more difficult. First, we replace the turbulent field in Eq. (\ref{peprwith4thorder})
by a {\it Fourier representation}
\be
\delta B_x \left( \vec{x} \right) = \int d^3 k \; \delta B_x \left( \vec{k} \right) e^{i \vec{k} \cdot \vec{x}}.
\label{fourierb}
\ee
To proceed we employ Corrsin's independence hypothesis (see Corrsin 1959)
\bdm
& & \left< \delta B_m \left( \vec{k} \right) \delta B_n^{*} \left( \vec{k}' \right) e^{i \vec{k} \cdot \vec{x} (t) - i \vec{k}' \cdot \vec{x} (0)} \right> \nonumber\\
& \approx & \left< \delta B_m \left( \vec{k} \right) \delta B_n^{*} \left( \vec{k}' \right) \right> \left< e^{i \vec{k} \cdot \vec{x} (t) - i \vec{k}' \cdot \vec{x} (0)} \right>
\label{Corrsin}
\edm
as well as the assumption of homogeneous turbulence
\be
\left< \delta B_m \left( \vec{k} \right) \delta B_n^{*} \left( \vec{k}' \right) \right>
= P_{mn} \left( \vec{k} \right) \delta \left( \vec{k} - \vec{k}' \right)
\label{homogturb}
\ee
where we have used Dirac's delta. Furthermore, we have used the magnetic correlation tensor
\be
P_{mn} \left( \vec{k} \right) = \left< \delta B_m \left( \vec{k} \right) \delta B_n^{*} \left( \vec{k} \right) \right>.
\label{defPmn}
\ee
By combining Eqs. (\ref{peprwith4thorder})-(\ref{defPmn}), we derive
\bdm
\kappa_{\perp} & = & \frac{v^2}{3 B_0^2} \int_{0}^{\infty} d t \; e^{-v t / \lambda_{\parallel}} \nonumber\\
& \times & \int d^3 k \; P_{xx} \left( \vec{k} \right) \left< e^{i \vec{k} \cdot \Delta \vec{x}} \right>
\label{anotherstep}
\edm
with $\Delta \vec{x} (t) = \vec{x} (t) - \vec{x} (0)$. To continue we combine Eq. (\ref{anotherstep}) with an Gaussian
distribution with vanishing mean. In this case the three-dimensional particle distribution function has the form
\bdm
f \left( \vec{x}, t \right) & = & \frac{1}{\sqrt{2 \pi \langle \left( \Delta x \right)^2 \rangle}}
\frac{1}{\sqrt{2 \pi \langle \left( \Delta y \right)^2 \rangle}} \frac{1}{\sqrt{2 \pi \langle \left( \Delta z \right)^2 \rangle}} \nonumber\\
& \times & e^{-\frac{x^2}{2 \left< \left( \Delta x \right)^2 \right>}} e^{-\frac{y^2}{2 \left< \left( \Delta y \right)^2 \right>}}
e^{-\frac{z^2}{2 \left< \left( \Delta z \right)^2 \right>}}.
\edm
For the axi-symmetric case this form gives the following characteristic function
\be
\left< e^{i \vec{k} \cdot \Delta \vec{x}} \right>
= e^{-\frac{1}{2} \left< \left( \Delta z \right)^2 \right> k_{\parallel}^2 -\frac{1}{2} \left< \left( \Delta x \right)^2 \right> k_{\perp}^2}
\label{characteristic}
\ee
where we have used cylindrical coordinates for the wave vector. Those are related to Cartesian coordinates via
\bdm
k_{\parallel} & = & k_{z}, \nonumber\\
k_{\perp} & = & \sqrt{k_{x}^2+k_{y}^2}, \nonumber\\
\Psi & = & \textnormal{arccot}(k_{x} / k_{y});
\label{cylindrical}
\edm
Furthermore, we assume that the particle motion along and across the mean magnetic field is diffusive and, therefore,
$\langle (\Delta z)^2 \rangle = 2 t \kappa_{\parallel}$ and $\langle (\Delta x)^2 \rangle = 2 t \kappa_{\perp}$.
We like to emphasize that for slab turbulence we have $k_{\perp} = 0$ in Eq. (\ref{characteristic}) and, thus, no assumption
has to be made for the perpendicular motion of the particle as long as slab turbulence is considered. By combining
this set of approximations and assumptions we can derive from Eq. (\ref{anotherstep})
\be
\kappa_{\perp} = \frac{a^2 v^2}{3 B_0^2} \int d^3 k \;
\frac{P_{xx} (\vec{k})}{v/\lambda_{\parallel} + \kappa_{\perp} k_{\perp}^2 + \kappa_{\parallel} k_{\parallel}^2}
\label{orignlgc}
\ee
which is a non-linear integral equation for $\kappa_{\perp}$. Here we have also introduced the {\it correction factor} $a^2$
as it was done in Matthaeus et al. (2003). In the latter paper it was suggested that $a^2 = 1/3$. Originally this parameter
was introduced in the equation of motion (\ref{eqofmotion}) but it was shown in Qin \& Shalchi (2016) that Eq. (\ref{eqofmotion})
in indeed valid as it is. Below the reader can find a more detailed discussion of this matter.
\subsection{The extended NLGC theory}
One can easily show that for slab turbulence Eq. (\ref{orignlgc}) provides a finite diffusion coefficient corresponding to normal
or {\it Markovian} diffusion. However, it is well-known that perpendicular transport in slab turbulence is subdiffusive (see, e.g.,
Qin et al. 2002a). Therefore, the ENLGC theory was developed in Shalchi (2005) and Shalchi (2006). In the following we present the
latter approach which was developed for two-component turbulence and cannot be used for any full three-dimensional turbulence model.

Eq. (\ref{eqofmotion}) with the Fourier representation (\ref{fourierb}) can be written as
\be
v_x = \frac{1}{B_0} \int d^3 k \; \delta B_x \left( \vec{k} \right) v_z e^{i \vec{k} \cdot \vec{x}}.
\label{formforvx}
\ee
In the slab model we have by definition $\delta B_x (\vec{x}) = \delta B_x (z)$ meaning that the turbulent field depends
only on the coordinate along the mean field. For pure slab turbulence we can use $\vec{k} \cdot \vec{x} = k_z z$ in
Eq. (\ref{formforvx}) and, therefore, we can write
\be
\frac{d}{d t} \Delta x = \frac{1}{B_0} \int d^3 k \; \frac{1}{i k_z} \delta B_x \left( \vec{k} \right) \frac{d}{d t} e^{i k_z z}.
\ee
The latter equation can easily be integrated over time to find
\be
\Delta x = \frac{1}{B_0} \int d^3 k \; \frac{1}{i k_z} \delta B_x \left( \vec{k} \right) \left[ e^{i k_z z(t)} - 1 \right]
\ee
where we have used $\Delta x (t) = x (t) - x (0)$ as well as $z (0) = 0$. The ensemble averaged square of this formula is
\bdm
\left< \left( \Delta x \right)^2 \right> & = & \frac{1}{B_0^2} \int d^3 k \; k_z^{-2} P_{xx} \left( \vec{k} \right) \nonumber\\
& \times & \left[ 2 - \left< e^{i k_z z(t)} \right> - \left< e^{-i k_z z(t)} \right> \right]
\label{msdslab1}
\edm
where we have employed again Eqs. (\ref{Corrsin}) - (\ref{defPmn}). Usually we are interested in the late time limit of the transport.
In this case, and by assuming that the motion in the parallel direction is diffusive in that limit, we can employ the characteristic
function of the diffusion equation
\be
\left< e^{\pm i k_z z(t)} \right> = e^{- \kappa_{\parallel} k_{\parallel}^2 t}.
\ee
It has to be emphasized that we only assumed that parallel transport is diffusive. No assumption was made concerning the perpendicular
motion. Therefore, Eq. (\ref{msdslab1}) can be written as
\be
\left< \left( \Delta x \right)^2 \right> = \frac{2}{B_0^2} \int d^3 k \; k_{\parallel}^{-2} P_{xx} \left( \vec{k} \right)
\left( 1 - e^{- \kappa_{\parallel} k_{\parallel}^2 t} \right).
\label{msdslab2}
\ee
The tensor components of the slab modes have the form
\be
P_{mn}^{slab} (\vec{k}) = g^{slab}(k_{\parallel}) \frac{\delta (k_{\perp})}{k_{\perp}} \delta_{mn},
\label{Plmslab}
\ee
with $m,n=x,y$. Here we have used the {\it Kronecker delta} $\delta_{mn}$ and the {\it Dirac delta} $\delta (k_{\perp})$, respectively.
The other tensor components are zero due to the solenoidal constraint. Furthermore, we have used the turbulence spectrum of the slab
modes $g^{slab}(k_{\parallel})$.

If we combine Eqs. (\ref{msdslab2}) and (\ref{Plmslab}), we derive
\be
\left< \left( \Delta x \right)^2 \right> = \frac{4 \pi \kappa_{\parallel}}{B_0^2} \int_{-\infty}^{+\infty} d k_{\parallel} \; g^{slab}(k_{\parallel})
\frac{1 - e^{- \kappa_{\parallel} k_{\parallel}^2 t}}{\kappa_{\parallel} k_{\parallel}^2}.
\label{msdslab3}
\ee
The fraction in this integral has the following property: If we consider the limit $t \rightarrow \infty$ the exponential goes to zero and the fraction
is finite as long as $k_{\parallel} \neq 0$. If $k_{\parallel} = 0$, however, the fraction is directly proportional to $t \rightarrow \infty$.
Therefore, the main contribution to the integral comes from very small wavenumbers $k_{\parallel}$. Thus we can write in the limit of late times
\be
\left< \left( \Delta x \right)^2 \right> \approx \frac{4 \pi \kappa_{\parallel}}{B_0^2} g^{slab}(k_{\parallel} = 0) \int_{-\infty}^{+\infty} d k_{\parallel} \; 
\frac{1 - e^{- \kappa_{\parallel} k_z^2 t}}{\kappa_{\parallel} k_{\parallel}^2}.
\label{msdslab4}
\ee
The remaining integral yields $2 \sqrt{\pi t / \kappa_{\parallel}}$ and, therefore, we obtain
\be
\left< \left( \Delta x \right)^2 \right> = \frac{8 \pi}{B_0^2} g^{slab}(k_{\parallel} = 0) \sqrt{\pi \kappa_{\parallel} t}.
\label{msdslab5}
\ee
The field line diffusion coefficient for slab turbulence is given by\footnote{For magnetostatic slab turbulence the theory of field line
random walk is exact. A field line diffusion coefficient $\kappa_{FL}$ is defined via the mean square displacements of magnetic field
lines $\langle ( \Delta x )^2 \rangle = 2 \kappa_{FL} \Delta z$ and, therefore, $\kappa_{FL}$ has length dimensions.} (see, e.g., Shalchi 2009)
\be
\kappa_{FL} = \frac{2 \pi^2}{B_0^2} g^{slab}(k_{\parallel} = 0)
\label{kappaFLwithg}
\ee
and, thus, Eq. (\ref{msdslab5}) can be written as
\be
\left< \left( \Delta x \right)^2 \right> = 4 \kappa_{FL} \sqrt{\frac{\kappa_{\parallel} t}{\pi}}.
\label{msdslab6}
\ee
For the spectrum of the slab modes we employ (see, e.g., Bieber et al. 1994)
\be
g^{slab} (k_{\parallel}) = \frac{C(s)}{2 \pi} \delta B_{slab}^2 l_{slab} \frac{1}{\left[ 1 + (k_{\parallel} l_{slab})^2 \right]^{s/2}}
\label{spectrum}
\ee
with the normalization function
\be
C(s) = \frac{\Gamma \left( \frac{s}{2} \right)}{2 \sqrt{\pi} \Gamma \left( \frac{s-1}{2} \right)}
\ee
where $\Gamma (z)$ is the {\it Gamma function}. Above we have used the magnetic field strength associated with the slab modes $\delta B_{slab}$, the slab
bendover scale $l_{slab}$, and the inertial range spectral index $s$. For this spectrum the field line diffusion coefficient (\ref{kappaFLwithg}) becomes
\be
\kappa_{FL} = \pi C(s) l_{slab} \frac{\delta B_{slab}^2}{B_0^2}.
\label{kappaFL}
\ee
From Eq. (\ref{msdslab6}) we can see that the mean square displacement increases linearly with $\sqrt{t}$ corresponding to subdiffusion.
In the literature this type of transport is usually called compound diffusion (see, e.g., K\'ota \& Jokipii 2000, Webb et al. 2006,
and Shalchi \& Kourakis 2007). Subdiffusion or compound diffusion was also discussed in the work of Getmantsev (1963), Fisk et al. (1973),
and Chuvilgin \& Ptuskin (1993). Furthermore, it was shown via test-particle simulations, that this type of transport can indeed be found
in slab turbulence (see, e.g., Qin et al. 2002a).

In the slab/2D composite model we assume that the magnetic field is given
by $\delta \vec{B} \left( \vec{x} \right) = \delta \vec{B}_{slab} \left( z \right) + \delta \vec{B}_{2D} \left( x, y \right)$ and we assume
that the two components are uncorrelated meaning that$\left< \delta B_{i,slab} \left( z \right) \delta B_{j,2D} \left( x, y \right) \right> = 0$.
It was shown before that if a two-dimensional component is added, diffusion is recovered (see, e.g., Qin et al. 2002b). Therefore, we assume
the following form for the mean square displacement\footnote{This choice for the mean square displacement is somewhat ad-hoc. Webb et al. (2009)
provided a Chapman-Kolmogorov approach to particle transport perpendicular to the mean field, in which the distribution function for magnetic field
diffusion was a Gaussian. A separate propagator for particle transport perpendicular and parallel with respect to the mean field was used. The mean
square displacement for particle transport across the mean field was given by Eq. (2.73) of that paper. This is similar to the form used here,
but there is a more complicated dependence of $\langle \left( \Delta x \right)^2 \rangle$ on time $t$ at intermediate times.}
\be
\left< \left( \Delta x \right)^2 \right> = 2 \alpha \sqrt{t} + 2 \kappa_{\perp} t
\label{msdform}
\ee
where, according to Eq. (\ref{msdslab6}),
\be
\alpha = 2 \kappa_{FL} \sqrt{\frac{\kappa_{\parallel}}{\pi}}.
\label{defalpha}
\ee
Very easily one can see that for $t \rightarrow \infty$, the (subdiffusive) slab contribution can be neglected compared to the second (diffusive)
contribution and the diffusion coefficient $\kappa_{\perp}$ depends only on the properties of the two-dimensional modes. Below we
will show that there can be an implicit contribution due to slab modes.

Within ENLGC theory, the slab contribution is calculated as described above, and the two-dimensional contribution as within the original
NLGC theory. Therefore, withing ENLGC theory the perpendicular diffusion coefficient in two-component turbulence is given by
\be
\kappa_{\perp} = \frac{v^2}{3 B_0^2} \int d^3 k \;
\frac{P_{xx}^{2D} (\vec{k})}{v/\lambda_{\parallel} + \kappa_{\perp} k_{\perp}^2}.
\label{extnlgc}
\ee
One can easily see that for pure two-dimensional turbulence Eqs. (\ref{orignlgc}) and (\ref{extnlgc}) are equivalent. As soon
as a slab contribution is added, however, both theories provide different results. Like in the original NLGC theory, one could
incorporate a correction factor $a^2$ but this is not done here.
\subsection{The UNLT theory}
According to Shalchi (2010), the original NLGC theory fails in general because approximation (\ref{4thorderapprox}) is not valid. This is in
particular the case for slab and small Kubo number turbulence. The latter statement was confirmed numerically in Qin \& Shalchi (2016).
Based on the CR Fokker-Planck equation, Shalchi (2010) developed a non-linear theory for perpendicular diffusion which does no longer
require the usage of approximation (\ref{4thorderapprox}). The following non-linear integral equation has been found after lengthy algebra
\be
\kappa_{\perp} = \frac{a^2 v^2}{3 B_0^2} \int d^3 k \; \frac{P_{xx} (\vec{k})}{v/\lambda_{\parallel} + (4/3) \kappa_{\perp} k_{\perp}^2 + F(k_{\parallel},k_{\perp})}
\label{UNLT}
\ee
with
\be
F \left( k_{\parallel}, k_{\perp} \right) = \frac{v^2 k_{\parallel}^2}{3 \kappa_{\perp} k_{\perp}^2}.
\ee
One can easily see that for two-dimensional turbulence Eq. (\ref{UNLT}) agrees with Eqs. (\ref{orignlgc}) and (\ref{extnlgc}) apart from
the factor $4/3$. For slab turbulence we find $\kappa_{\perp} = 0$ and, therefore, UNLT and ENLGC theories agree with each other but not
with the original NLGC theory. Whereas ENLGC theory can only be used for two-component turbulence, UNLT theory should be valid for full
three-dimensional turbulence also. From Eq. (\ref{UNLT}) one can easily derive the Matthaeus et al. (1995) theory for field line random
walk by considering the limit $\lambda_{\parallel} \rightarrow \infty$. Different asymptotic limits of Eq. (\ref{UNLT}) have been derived
and discussed in Shalchi (2015).
\section{Implicit contribution of slab modes}
Within the extended NLGC theory the mean square displacement is given by Eq. (\ref{msdform}). Obviously there is a subdiffusive
contribution from the slab modes and a diffusive contribution from the two-dimensional modes. In the limit $t \rightarrow \infty$,
however, only the latter contribution remains.

In previous analytical theories such as the ones described in Sect. 2, one usually employs $\langle ( \Delta x )^2 \rangle = 2 \kappa_{\perp} t$
in the characteristic function. The idea of the current paper is to use Eq. (\ref{msdform}) instead of the diffusion approximation.
For two-dimensional turbulence Eq. (\ref{anotherstep}) with (\ref{characteristic}) and (\ref{msdform}) becomes
\bdm
\kappa_{\perp} & = & \frac{v^2}{3 B_0^2} \int d^3 k \; P_{xx}^{2D} \left( \vec{k} \right) \nonumber\\
& \times & \int_{0}^{\infty} d t \; e^{- \left( v/\lambda_{\parallel} + \kappa_{\perp} k_{\perp}^2 \right) t - \alpha k_{\perp}^2 \sqrt{t}}.
\label{nlgc1}
\edm
It has to be emphasized that the latter formula is only valid for two-dimensional turbulence and cannot be used for full three-dimensional
turbulence. Furthermore, the form (\ref{msdform}) is only valid in the late time limit. For earlier times, for instance, one expects
a ballistic motion of particles and for intermediate times there could even be a diffusive contribution of the slab modes (see, e.g., Jokipii \& Parker 1969,
Shalchi 2008, Webb et al. 2009, and Ruffolo et al. 2012 for more details). The model used here is based on the assumption that only
late times contribute to the perpendicular diffusion coefficient.

The time integral in Eq. (\ref{nlgc1}) is solved by (see, e.g., Gradshteyn \& Ryzhik 2000)
\be
\int_{0}^{\infty} d t \; e^{- A \sqrt{t} - B t} = \frac{1}{B} K \left( \xi \right)
\ee
where we have used the parameters
\be
A = \alpha k_{\perp}^2,
\ee
\be
B = v / \lambda_{\parallel} + \kappa_{\perp} k_{\perp}^2,
\ee
and
\be
\xi = \frac{A}{2 \sqrt{B}},
\label{definexi}
\ee
as well as the function
\be
K \left( \xi \right) = 1 - \sqrt{\pi} \xi e^{\xi^2} \textnormal{erfc} \left( \xi \right).
\label{definefuncK}
\ee
Here we have also used the {\it complementary error function}. Therewith, Eq. (\ref{nlgc1}) becomes
\be
\kappa_{\perp} = \frac{v^2}{3 B_0^2} \int d^3 k \; \frac{P_{xx}^{2D} (\vec{k})}{v/\lambda_{\parallel} + \kappa_{\perp} k_{\perp}^2} K \left( \xi \right).
\label{nlgc2}
\ee
The parameter $\xi$ was defined in Eq. (\ref{definexi}) and becomes in our case
\bdm
\xi & = & \frac{\kappa_{FL} k_{\perp}^2 \sqrt{\kappa_{\parallel} / \pi}}{\sqrt{ v / \lambda_{\parallel} + \kappa_{\perp} k_{\perp}^2}} \nonumber\\
& = & \frac{1}{\sqrt{3 \pi}} \frac{\kappa_{FL} \lambda_{\parallel} k_{\perp}^2}{\sqrt{ 1 + \lambda_{\parallel} \lambda_{\perp} k_{\perp}^2 /3}}
\label{formofxi}
\edm
where we have used Eq. (\ref{defalpha}) and we have replaced the diffusion coefficients by the corresponding mean free paths\footnote{The mean free paths
are related to the corresponding diffusion coefficients via $\lambda_{\parallel} = 3 \kappa_{\parallel} / v$ and $\lambda_{\perp} = 3 \kappa_{\perp} / v$.}.
If we employ Eq. (\ref{kappaFL}) in order to replace the field line diffusion coefficient of the slab modes, we can write
\be
\xi = \sqrt{\frac{\pi}{3}} C(s) \frac{l_{slab} \lambda_{\parallel} k_{\perp}^2}{\sqrt{ 1 + \lambda_{\parallel} \lambda_{\perp} k_{\perp}^2 /3}} \frac{\delta B_{slab}^2}{B_0^2}.
\ee
If we replace the diffusion coefficients by the corresponding mean free paths, Eq. (\ref{nlgc2}) can be written as
\be
\lambda_{\perp} = \frac{\lambda_{\parallel}}{B_0^2} \int d^3 k \;
\frac{P_{xx}^{2D} (\vec{k})}{1 + \lambda_{\parallel} \lambda_{\perp} k_{\perp}^2 / 3} K \left( \xi \right).
\label{nlgc3}
\ee
The two-dimensional turbulence model is defined via
\be
P_{mn} (\vec{k}) = g^{2D} (k_{\perp}) \frac{\delta (k_{\parallel})}{k_{\perp}} \left[ \delta_{mn} - \frac{k_m k_n}{k_{\perp}^2} \right]
\label{def2d}
\ee
where we have used the two-dimensional turbulence spectrum $g^{2D} (k_{\perp})$ which will be discussed below. By combining Eqs. (\ref{nlgc3})
and (\ref{def2d}) we obtain
\be
\lambda_{\perp} = \frac{\pi \lambda_{\parallel}}{B_0^2} \int_{0}^{\infty} d k_{\perp} \; \frac{g^{2D} (k_{\perp})}{1 + \lambda_{\parallel} \lambda_{\perp} k_{\perp}^2 /3} K \left( \xi \right).
\label{withg2d}
\ee
Shalchi \& Weinhorst (2009) proposed the following form for the spectrum of the two-dimensional modes
\bdm
g^{2D}(k_{\perp}) & = & \frac{2 D(s, q)}{\pi} \delta B_{2D}^2 l_{2D} \nonumber\\
& \times & \frac{(k_{\perp} l_{2D})^{q}}{\left[ 1 + (k_{\perp} l_{2D})^2 \right]^{(s+q)/2}}
\label{2dspec}
\edm
with the normalization function
\be
D(s, q) = \frac{\Gamma \left( \frac{s+q}{2} \right)}{2 \Gamma \left( \frac{s-1}{2} \right) \Gamma \left( \frac{q+1}{2} \right)}.
\label{normalD}
\ee
The parameters used in the spectrum are the inertial range spectral index $s$, the energy range spectral index $q$, and the bendover scale
of the two-dimensional modes $l_{2D}$. With this spectrum, and by employing the integral transformation $x = l_{2D} k_{\perp}$,
Eq. (\ref{withg2d}) becomes
\bdm
\lambda_{\perp} & = & 2 D(s, q) \lambda_{\parallel} \frac{\delta B_{2D}^2}{B_0^2} \nonumber\\
& \times & \int_{0}^{\infty} d x \; \frac{x^q}{\left( 1 + x^2 \right)^{(s+q)/2}} \frac{K (\xi)}{1 + \frac{\lambda_{\parallel} \lambda_{\perp}}{3 l_{2D}^2} x^2}.
\label{finalnewnlgc}
\edm
In Sect. 5 we shall evaluate Eq. (\ref{finalnewnlgc}) for different parameter values and compare our findings with diffusion coefficients
obtained from the other theories.
\section{Further analytical considerations}
The important result of the current paper is Eq. (\ref{nlgc2}). Therein the function $K (\xi)$ is used which is defined via Eq. (\ref{definefuncK}).
It shouldn't be problematic to incorporate this function in numerical codes used to evaluate Eq. (\ref{nlgc2}). In some cases, however, a further
analytical simplification could be convenient. This is done in the following.

\begin{figure}
\includegraphics[scale=0.4]{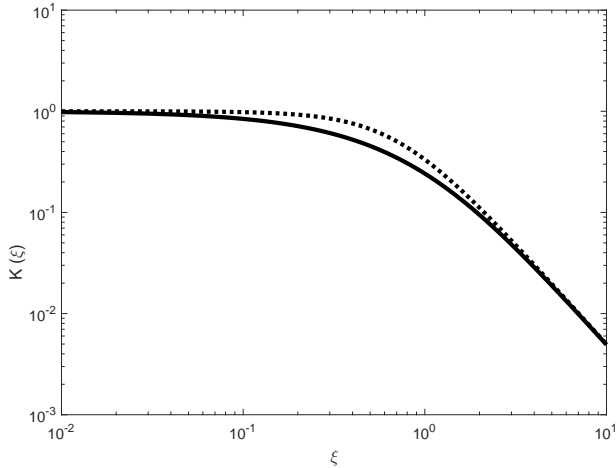}
\caption{The function $K(\xi)$. We show the exact analytical form given by Eq. (\ref{definefuncK}) (solid line) as well as
approximation (\ref{approxK}) (dotted line).}
\label{funcK}
\end{figure}

The complementary error function has the following asymptotic limits (see, e.g., Abramowitz \& Stegun 1974)
\be
\textnormal{erfc} \left( \xi \ll 1 \right) \approx 1 - \frac{2}{\sqrt{\pi}} \xi
\ee
and
\be
\textnormal{erfc} \left( \xi \gg 1 \right) \approx \frac{1}{\sqrt{\pi} \xi} e^{- \xi^2} \left( 1 - \frac{1}{2 \xi^2} \right).
\ee
Therefore,we find
\be
K \left( \xi \ll 1 \right) \approx 1 - \sqrt{\pi} \xi
\ee
and
\be
K \left( \xi \gg 1 \right) \approx \frac{1}{2 \xi^2}.
\label{largexi}
\ee
We can easily see that for increasing $\xi$, we find a reduction of the perpendicular diffusion coefficient. In order to combine our findings
we use the following approximation
\be
K \left( \xi \right) \approx \frac{1}{1 + 2 \xi^2}
\label{approxK}
\ee
so that $K (\xi \ll 1) \approx 1$ and for $\xi \gg 1$ we recover Eq. (\ref{largexi}). If Fig. \ref{funcK}, we compare the exact
form (\ref{definefuncK}) with approximation Eq. (\ref{approxK}).

If approximation (\ref{approxK}) is combined with Eq. (\ref{nlgc2}), and if we use Eq. (\ref{formofxi}) we find the following integral equation
\be
\kappa_{\perp} = \frac{v^2}{3 B_0^2} \int d^3 k \;
\frac{P_{xx}^{2D} (\vec{k})}{v/\lambda_{\parallel} + \kappa_{\perp} k_{\perp}^2 + 2 \kappa_{\parallel} \kappa_{FL}^2 k_{\perp}^4 / \pi}.
\label{finalapprox}
\ee
The latter formula has some similarity with the integral equations discussed in Sect. 2 (see, e.g., Eqs. (\ref{orignlgc}), (\ref{extnlgc}),
and (\ref{UNLT})). We like to emphasize that Eq. (\ref{finalapprox}) is only valid for slab/2D turbulence and cannot be used for other turbulence
models such as full three-dimensional models. The third term in the denominator of Eq. (\ref{finalapprox}) contains the field line diffusion
coefficient $\kappa_{FL}$. It has to be pointed out that this is the field line diffusion coefficient associated with the slab modes as given
by Eq. (\ref{kappaFL}) and not the total field line diffusion coefficient which would also contain a contribution of the two-dimensional modes.

For numerical investigations it is useful to rewrite Eq. (\ref{finalapprox}) as
\bdm
\lambda_{\perp} & = & 2 D(s, q) \lambda_{\parallel} \frac{\delta B_{2D}^2}{B_0^2} \nonumber\\
& \times & \int_{0}^{\infty} d x \; \frac{x^q}{\left( 1 + x^2 \right)^{(s+q)/2}} \frac{1}{1 + \frac{\lambda_{\parallel} \lambda_{\perp}}{3 l_{2D}^2} x^2 + \gamma x^4}
\label{finalapproxmfp}
\edm
where we have used the parameter
\bdm
\gamma & = & \frac{2}{3 \pi} \frac{\lambda_{\parallel}^2 \kappa_{FL}^2}{l_{2D}^4} \nonumber\\
& = &\frac{2 \pi}{3} \left[ C \left( s \right) \right]^2 \frac{\lambda_{\parallel}^2}{l_{2D}^2} \frac{l_{slab}^2}{l_{2D}^2} \frac{\delta B_{slab}^4}{B_0^4}
\label{parametergamma}
\edm
and spectrum (\ref{2dspec}). Eq. (\ref{finalapproxmfp}) with (\ref{parametergamma}) is also used in Sect. 5 to compute the perpendicular
mean free path and the results are compared with other theoretical results as well.
\section{Results}
In the current section we compute the perpendicular mean free path by employing the original NLGC theory of Matthaeus et al. (2003),
the extended NLGC theory of Shalchi (2006), and we use the modified theory developed in the current paper by using different approximations
for the function $K (\xi)$. In all cases we calculate the perpendicular mean free path as a function of the parallel mean free path.
In all cases we have set $\delta B_{slab}^2 = 0.2 B_0^2$ and $\delta B_{2D}^2 = 0.8 B_0^2$ as originally suggested in Bieber et al. (1994).
\subsection{The case $q=0$ and $l_{2D} = 0.1 l_{slab}$}
\begin{figure}
\includegraphics[scale=0.4]{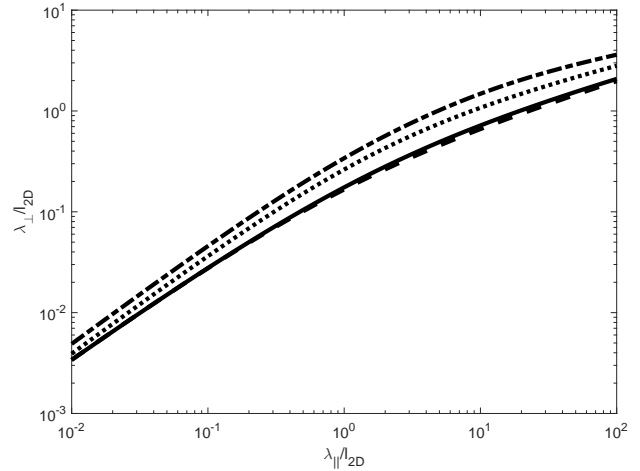}
\caption{The perpendicular mean free path $\lambda_{\perp} / l_{2D}$ versus the parallel mean free path $\lambda_{\parallel} / l_{2D}$
for $q=0$ and $l_{2D} = 0.1 l_{slab}$. Shown are the results from the original NLGC theory (dash-dotted line), the extended NLGC theory (dotted line),
and the modified approach developed in the current paper. The latter theory was evaluated by using the exact form for the function $K (\xi)$ (solid line)
and approximation (\ref{approxK}) (dashed line).}
\label{q0l10}
\end{figure}

The first set of parameter values is based on those used in Matthaeus et al. (2003). In the latter paper the original NLGC
theory was compared with test-particle simulations. The best agreement was achieved by setting $a^2 = 1/3$. In Fig. \ref{q0l10}
we show the original NLGC theory for $a^2 = 1$, the extended NLGC theory as well as our new results. We can clearly see that the
extended NLGC result is smaller than the original NLGC result because there is no contribution from the slab modes. Furthermore,
the implicit contribution from the slab modes reduces the perpendicular mean free path further. This is in particular the case for
long parallel mean free paths corresponding to higher particle rigidities/energies. In this case the perpendicular mean free path
is reduced by about a factor two compared to the original NLGC result. This finding can explain the value $a^2 \approx 1/3$ suggested
by Matthaeus et al. (2003). We also computed the perpendicular mean free path for $q=0$ and $l_{2D} = l_{slab}$ but did not observe
a significant effect.

\subsection{The case $q=1.5$ and $l_{2D} = 0.1 l_{slab}$}
\begin{figure}
\includegraphics[scale=0.4]{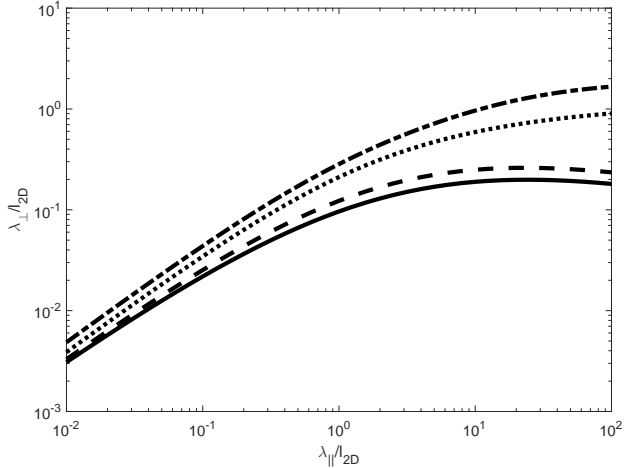}
\caption{Caption is as in Fig. \ref{q0l10} but here we have used $q=1.5$ and $l_{2D} = 0.1 l_{slab}$ to explore the influence of the
energy range spectral index $q$.}
\label{q15l10}
\end{figure}

Matthaeus et al. (2007) suggested that the spectrum of the two-dimensional modes is not constant at large scales corresponding to the
energy range. Therefore, we set $q=1.5$ and compute the perpendicular mean free path as it was done above. Our findings are shown
in Fig. \ref{q15l10}. Clearly we can observe a significant difference between the different theories. For the case considered here,
the implicit contribution of the slab modes reduces the perpendicular mean free path by about a factor $10$ if compared with the
original NLGC theory. However, this is only the case for very long parallel mean free paths corresponding to very high particle
energies. We can also see that approximation (\ref{approxK}) works well.

\subsection{The case $q=3$ and $l_{2D} = 0.1 l_{slab}$}
\begin{figure}
\includegraphics[scale=0.4]{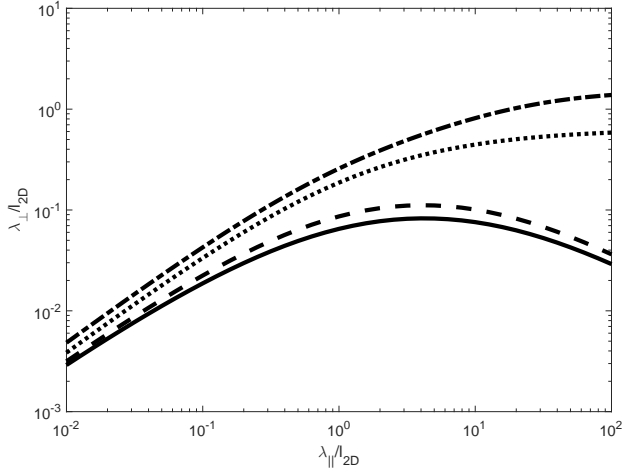}
\caption{Caption is as in Fig. \ref{q0l10} but here we have used $q=3$ and $l_{2D} = 0.1 l_{slab}$ to explore the influence of the
energy range spectral index $q$.}
\label{q3l10}
\end{figure}

As shown above, the energy range spectral index seems to be important if the implicit slab contribution is taken into account. Therefore,
we further change the parameter $q$. Our findings for $q=3$ are visualized in Fig. \ref{q3l10}. Again we find a significant difference
between the different theories. Now the influence of the implicit slab contribution is even larger and, thus, we conclude that for
increasing $q$ we find a stronger effect.

\subsection{The case $q=3$ and $l_{2D} = l_{slab}$}
\begin{figure}
\includegraphics[scale=0.4]{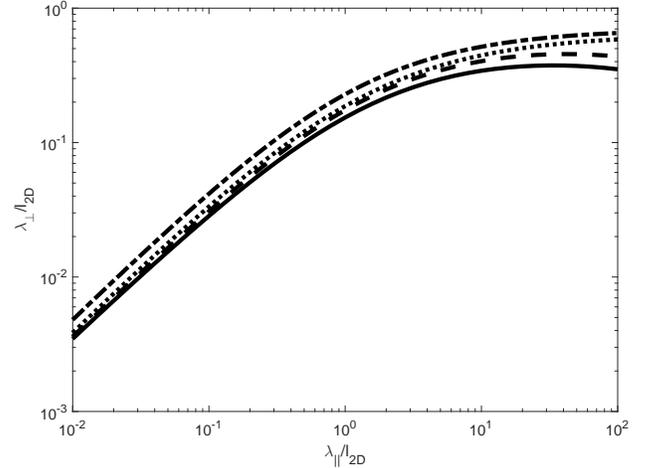}
\caption{Caption is as in Fig. \ref{q0l10} but here we have used $q=3$ and $l_{2D} = l_{slab}$ to explore the influence of the bendover
scales.}
\label{q3l1}
\end{figure}

A further important parameter in the theory of perpendicular diffusion is the scale ratio $l_{2D} / l_{slab}$. Above we have considered the case
of $l_{2D} = 0.1 l_{slab}$ as it was used in Matthaeus et al. (2003). In the current paragraph we assume that the two bendover scales are equal.
Our findings are shown in Fig. \ref{q3l1}. Clearly we can see that now the discrepancies between the different theories are much smaller. However,
there is still a factor $2$ or even $3$ between the different theoretical results. Again the observed effect can explain the value $a^2 = 1/3$ assumed
in Matthaeus et al. (2003). We also made calculations for smaller values of the magnetic field ratio $\delta B / B_0$ to find that the effect coming
from the implicit slab contribution is weaker.

\section{Summary and Conclusion}
In the current paper we have revisited the problem of perpendicular diffusion of energetic particles in two-component turbulence.
Whereas it was shown before that the slab modes do not explicitly contribute to the perpendicular diffusion coefficient, we explored
the implicit contribution in the current paper. We have derived the modified non-linear integral equation (\ref{nlgc2}) which can be
approximated by Eq. (\ref{finalapprox}). This modification should provide another improvement compared to the original NLGC theory
developed in Matthaeus et al. (2003) and the extended NLGC theory of Shalchi (2006). Compared to earlier versions of the NLGC theory,
the modified equations (\ref{nlgc2}) or (\ref{finalapprox}) are not more difficult to evaluate numerically.

In Figs. \ref{q0l10}-\ref{q3l1} we have shown the perpendicular mean free path versus the parallel mean free path for different values
of the scale ratio $l_{2D} / l_{slab}$ and different values of the energy range spectral index $q$. Both mean free paths are normalized
with respect to the two-dimensional bendover scale $l_{2D}$. It can easily be seen that, in general, the implicit slab contribution
reduces the perpendicular mean free path.

Matthaeus et al. (2003) used $q=0$ and $l_{2D} = 0.1 l_{slab}$ in their work and they compared the original NLGC theory with test-particle
simulations. They found a difference between analytical theory and simulations but this difference can be balanced out by using the correction
factor $a^2$ and by setting $a^2 = 1/3$. In the current paper a possible explanation for this value is provided. The implicit contribution
from the slab modes reduces the perpendicular mean free path as required to achieve agreement with simulations. The correction factor $a^2$
is no longer needed.

For $q=1.5$ and $q=3$, which is in agreement with the values suggested by Matthaeus et al. (2007), we find a stronger effect. In particular
for long parallel mean free paths, a strong reduction of the perpendicular mean free path can be observed.

Another parameter which influences the reduction discussed here, is the scale ratio $l_{2D} / l_{slab}$. If this ratio is small, a stronger
effect can be observed. For equal bendover scales, however, the perpendicular mean free path is only about a factor $2$ shorter
as the one computed by using the original NLGC theory.

A further theory for perpendicular diffusion was presented in Shalchi (2010) where the Unified Non-Linear Transport (UNLT) theory was
developed. In the following we discuss which theory has to be used for which case.
\begin{itemize}
\item Solar Wind turbulence is often approximated by a slab/2D composite model which is also known as two-component turbulence.
We suggest to use the extended NLGC theory with implicit slab contribution developed in the current paper for this specific turbulence
model. This theory is represented by Eq. (\ref{nlgc2}) which can be well approximated by Eq. (\ref{finalapprox}).
\item For full three-dimensional turbulence, the original NLGC theory, the extended theory and the approach developed in the current
paper cannot be used. For this case the UNLT theory represented by Eq. (\ref{UNLT}) should provide an accurate description of perpendicular
diffusion. In this case a critical parameter is the {\it Kubo number} (see Shalchi 2015 for more details).
\end{itemize}

The integral equation derived in the current paper should provide an accurate analytical description of perpendicular diffusion in
two-component turbulence. It is straightforward to include further effects such as dynamical turbulence. In this case another term
would occur in the denominator of Eq. (\ref{nlgc2}) which would be associated with the correlation time of the two-dimensional modes.
The situation is more complicated, however, if there is also a dynamical turbulence effect associated with the slab modes because
in this case the explicit contribution of the slab modes can be diffusive (see, e.g., Shalchi 2014).

The results obtained in the current paper, and in analytical theories for perpendicular diffusion in general, are relevant for
several applications:
\begin{itemize}
\item To understand the acceleration of particles due to turbulence (see, e.g., Lynn et al. 2014);
\item For solar modulation studies (see, e.g., Alania et al. 2013, Engelbrecht \& Burger 2013, Manuel et al. 2014, and Potgieter et al. 2014);
\item Particle acceleration at interplanetary shocks such as coronal mass ejection driven shocks (see, e.g., Li et al. 2012 and Wang et al. 2012);
\item To describe the motion of cosmic rays in our own and in external galaxies (see, e.g., Buffie et al. 2013, Berkhuijsen et al. 2013, Heesen et al. 2014);
\item To describe diffusive shock acceleration at interstellar shocks (see, e.g., Ferrand et al. 2014);
\end{itemize}
In particular for diffusive shock acceleration at supernova shock waves, the fact that the perpendicular mean free path becomes
rigidity independent in the high energy limit, can help to explain the cosmic ray spectrum (see Ferrand et al. 2014 for more details).
In the current paper we have shown that the perpendicular diffusion coefficient can even decrease with increasing rigidity in the high
energy regime. To incorporate this effect in simulations of diffusive shock acceleration at interplanetary and interstellar shock waves
could be important and should be subject of future work.
\begin{acknowledgements}
{\it A. Shalchi acknowledges support by the Natural Sciences and Engineering Research Council (NSERC) of Canada.}
\end{acknowledgements}
{}


\begin{thebibliography}{}

\bibitem[Abramowitz \& Stegun(1974)]{abr74}
Abramowitz, M., \& Stegun, I. A. 1974, Handbook of Mathematical Functions (New York: Dover Publications)

\bibitem[Alania et al.(2013)]{alan13}
Alania, M. V., Wawrzynczak, A., Sdobnov, V. E., \& Kravtsova, M. V. 2013, SoPh, 286, 561

\bibitem[Berkhuijsen et al.(2013)]{berkhuijsen2013}
Berkhuijsen, E. M., Beck, R., \& Tabatabaei, F. S. 2013, MNRAS, 435, 1598

\bibitem[Bieber et al.(1994)]{bie94}
Bieber, J. W., Matthaeus, W. H., Smith, C. W., Wanner, W., Kallenrode, M.-B., \& Wibberenz, G. 1994, ApJ, 420, 294

\bibitem[Buffie et al.(2013)]{buffie13}
Buffie, K., Heesen, V., \& Shalchi, A. 2013, ApJ, 764, 37

\bibitem[Chuvilgin \& Ptuskin(1993)]{Chuvil93}
Chuvilgin, L. G., \& Ptuskin, V. S. 1993, A\&A, 279, 278

\bibitem[Corrsin(1959)]{cor59}
Corrsin, S. 1959, in Atmospheric Diffusion and Air Pollution, Adv. Geophys. 6, Progress report on some
turbulent diffusion research, ed. F. Frenkiel, \& P. Sheppard (New York: Academic)

\bibitem[Engelbrecht \& Burger(2013)]{engel2013}
Engelbrecht, N. E. \& Burger, R. A. 2013, ApJ, 779, 158

\bibitem[Ferrand et al.(2014)]{Ferrand14}
Ferrand, G., Danos R. J., Shalchi, A., Safi-Harb, S., Edmon, P., \& Mendygral, P. 2014, ApJ, 792, 133

\bibitem[Fisk et al.(1973)]{Fisk73}
Fisk, L. A., Ramaty, R., \& Lingenfelter, R. E. 1973, Proceedings of the 13th International Conference on Cosmic Rays,
held in Denver, Colorado, Volume 1 (OG Sessions), p. 367

\bibitem[Getmantsev(1963)]{Getman63}
Getmantsev, G. G. 1963, Soviet Astronomy, 6, 477

\bibitem[Gradshteyn \& Ryzhik(2000)]{GradshteynRyzhik00}
Gradshteyn, I .S., \& Ryzhik, I. M. 2000, Table of integrals, series, and products (New York: Academic Press)

\bibitem[Green(1951)]{green51}
Green, M. S. 1951, Journal of Chemical Physics, 19, 1036

\bibitem[Heesen et al.(2014)]{heesen14}
Heesen, V., Croston, J. H., Harwood, J. J., Hardcastle, M. J. \& Hota, A. 2014, MNRAS, 439, 1364

\bibitem[Jokipii(1966)]{jok66}
Jokipii, J. R. 1966, ApJ, 146, 480

\bibitem[Jokipii \& Parker(1969)]{jokpar69}
Jokipii, J. R., \& Parker, E. N. 1969, ApJ, 155, 777

\bibitem[K\'ota(2000)]{kota00}
K\'ota, J., \& Jokipii, J. R. 2000, ApJ, 531, 1067

\bibitem[Kubo(1957)]{kubo57}
Kubo, R. 1957, J. Phys. Soc. Jpn., 12, 570

\bibitem[Li et al.(2012)]{Li2012}
Li, G., Shalchi, A., Ao, X., Zank, G., \& Verkhoglyadova, O. P. 2012, AdSpR, 49, 1067

\bibitem[Lynn et al.(2014)]{lynn14}
Lynn, J. W., Quataert, E., Chandran, B. D. G., \& Parrish, I. J. 2014, ApJ, 791, 71

\bibitem[Manuel et al.(2014)]{manuel14}
Manuel, R., Ferreira, S. E. S., \& Potgieter, M. S. 2014, SoPh, 289, 2207

\bibitem[Matthaeus et al.(1995)]{matt95}
Matthaeus, M. W., Gray, P. C., Pontius Jr., D. H., \& Bieber, J. W. 1995, Phys. Rev. Lett., 75, 2136

\bibitem[Matthaeus et al.(2003)]{matt03}
Matthaeus, W. H., Qin, G., Bieber, J. W., \& Zank, G. P. 2003, ApJ, 590, L53

\bibitem[Matthaeus et al.(2007)]{matt07}
Matthaeus, W. H., Bieber, J. W., Ruffolo, D., Chuychai, P., \& Minnie, J. 2007, ApJ, 667, 956

\bibitem[Owens(1974)]{Owens74}
Owens, A. J. 1974, ApJ, 191, 235

\bibitem[Potgieter et al.(2014)]{potgieter2014}
Potgieter, M. S., Vos, E. E., Boezio, M., De Simone, N., Di Felice, V., \& Formato, V. 2014, SoPh, 289, 391

\bibitem[Qin et al.(2002a)]{qin02a}
Qin, G., Matthaeus, W. H., \& Bieber, J. W. 2002a, Geophys. Res. Lett., 29, 1048

\bibitem[Qin et al.(2002b)]{qin02b}
Qin, G., Matthaeus, W. H., \& Bieber, J. W. 2002b, ApJ,  578, L117

\bibitem[Qin \& Shalchi(2016)]{qinsha16}
Qin, G., \& Shalchi, A. 2016, ApJ, 823, 23

\bibitem[Ruffolo et al.(2012)]{ruffolo2012}
Ruffolo, D., Pianpanit, T., Matthaeus, W. H., \& Chuychai, P. 2012, ApJ, 747, L34

\bibitem[Schlickeiser(2002)]{Schlickeiser2002}
Schlickeiser R. 2002, Cosmic Ray Astrophysics (Berlin:Springer)

\bibitem[Shalchi(2005)]{shalchi05jgr}
Shalchi, A. 2005, JGR, 110, A09103

\bibitem[Shalchi(2006)]{shal06}
Shalchi, A. 2006, A\&A, 453, L43

\bibitem[Shalchi \& Kourakis(2007)]{shalkou2007}
Shalchi, A., \& Kourakis, I. 2007, A\&A, 470, 405

\bibitem[Shalchi(2008)]{shal08}
Shalchi, A. 2008, Plasma Physics and Controlled Fusion, 50, 055001

\bibitem[Shalchi(2009)]{shal09book}
Shalchi, A. 2009, Nonlinear Cosmic Ray Diffusion Theories, Astrophysics and Space Science Library, Vol. 362
(Berlin: Springer)

\bibitem[Shalchi \& Weinhorst(2009)]{Shalwei2009}
Shalchi, A., \& Weinhorst, B. 2009, AdSpR, 43, 1429

\bibitem[Shalchi(2010)]{shal2010}
Shalchi, A. 2010, ApJL, 720, L127

\bibitem[Shalchi(2011)]{shal2011}
Shalchi, A. 2011, AdSpR, 47, 1147

\bibitem[Shalchi(2014)]{shal2014}
Shalchi, A. 2014, ApJ, 780, 138

\bibitem[Shalchi \& Hussein(2014)]{shalhuss2014}
Shalchi, A., \& Hussein, M. 2014, ApJ, 794, 56

\bibitem[Shalchi(2015)]{shal2015}
Shalchi, A. 2015, PhPl, 22, 010704

\bibitem[Tautz \& Shalchi(2011)]{tautzshal2011}
Tautz, R. C., \& Shalchi, A. 2011, ApJ, 735, 92

\bibitem[Taylor(1922)]{taylor22}
Taylor, G. I. 1922, Proceedings of the London Mathematical Society, 20, 196

\bibitem[Wang et al.(2012)]{wang12}
Wang, Y., Qin, G., \& Zhang, M. 2012, ApJ, 752, 37

\bibitem[Webb et al.(2006)]{web06}
Webb, G. M., Zank, G. P., Kaghashvili, E. Kh., \& le Roux, J. A. 2006, ApJ, 651, 211

\bibitem[Webb et al.(2009)]{web09}
Webb, G. M., Kaghashvili, E. Kh, le Roux, J. A., Shalchi, A., Zank, G. P., \& Li, G. 2009,
Journal of Physics A: Mathematical and Theoretical, 42, 235502

\end{thebibliography}
\end{document}